\documentclass[aps,prl,twocolumn,showpacs,superscriptaddress,groupedaddress]{revtex4}  % for review and submission
\usepackage{graphicx}  % needed for figures
\usepackage{dcolumn}   % needed for some tables
\usepackage{bm}        % for math
\usepackage{amssymb}   % for math
\usepackage{amsmath}
\begin{document}

% the following line is for submission, including submission to the arXiv!!
%\hspace{5.2in} \mbox{Fermilab-Pub-04/xxx-E}

\title{Spin Orbit Magnetism and Unconventional Superconductivity}

\author{
Yi Zhang and Kevin S. Bedell\\
\textit{Department of Physics, Boston College, Chestnut Hill, Massachusetts 02467, USA}
}

\date{\today}

\begin{abstract}
We find an exotic spin excitation in a magnetically ordered system with spin orbit magnetism in 2D, where the order parameter has a net spin current and no net magnetization. Starting from a Fermi liquid theory, similar to that for a weak ferromagnet, we show that this excitation emerges from an exotic magnetic Fermi liquid state that is protected by a generalized Pomeranchuck condition. We derive the propagating mode using the Landau kinetic equation, and find that the dispersion of the mode has a $\sqrt q$ behavior in leading order in 2D. We find an instability toward superconductivity induced by this exotic mode, and a further analysis based on the forward scattering sum rule strongly suggests that this superconductivity has p-wave pairing symmetry. We perform similar studies in the 3D case, with a slightly different magnetic system and find that the mode leads to a Lifshitz-like instability most likely toward an inhomogeneous magnetic state in one of the phases.
\end{abstract}
\maketitle

The Landau Fermi liquid theory is a very successful theory in condensed matter physics. It provides a phenomenological framework for describing thermodynamics, transport and collective modes of itinerant fermionic systems. In the Landau theory, the interactions among quasi-particles are described by the Landau parameters $F_l^{s,a}$, where $l$ denotes the orbital angular momentum partial-wave channel, and $s,a$ denote spin symmetric and antisymmetric channels, respectively. It has been proved by Pomeranchuck that for the Fermi surface to be stable, the Landau parameters should satisfy the relation: $F_l^{s,a}>-(2l+1)$. Whenever the relation is violated, there will exist an instability of the Fermi surface known as a Pomeranchuck instability \cite{P}, such as the Stoner ferromagnetism when $F_0^a\rightarrow-1^+$, or phase separation when $F_0^s\rightarrow-1^+$. In 1959, Abrikosov and Dzyaloshinskii \cite{AD} developed a ferromagnetic Fermi liquid theory(FFLT) of itinerant ferromagnetism based on Landau Fermi liquid theroy, whose microscopic foundations were established later by Dzyaloshiskii and Kondratenko \cite{DK}. Further studies had been made of this state using a generalized Pomeranchuck instability based on the FFLT of Blagoev, Engelbrecht and Bedell \cite{BEB} and Bedell and Blagoev \cite{BB}.

Recently, Pomeranchuck instabilities in higher angular momentum partial-wave channels have been studied by many authors, such as the quantum nematic Fermi liquid phase as a result of an instability in the $F_2^s$ channel by Oganesyan, Kivelson and Fradkin \cite{OKF} and the so called $\alpha$ and $\beta$ phases the in $F_1^a$ channel by Wu, Zhang, et al. \cite{WZ,WSZ}. In these papers, mean-field theory is used based on the microscopic Hamiltonian to demonstrate the instabilities of the disordered phase and to classify the possible phases of the ordered state. The Goldstone modes are studied within the random-phase-approximation(RPA) approach.

In this paper, we use the traditional Fermi liquid theory, similar to the FFLT, in the weak magnetic limit, to study the generalized Pomeranchuck instability in the $F_1^a$ channel in a 2D system. Here we start from the state with an ordered phase, using the Landau kinetic equation to study the collective modes. In this symmetry broken phase we find an exotic collective mode. We further find a superconducting instability induced by this mode. We also carry out a similar calculation in a 3D system with a slightly different model, and find the mode leads to a Lifshitz-like instability toward an inhomogeneous magnetic state.

Similar to what was done in the weakly ferromagnetic system \cite{BEB}, we expand the deviation of the energy around the ordered ground state in the spirit of Landau up to second order in the deviations, $\delta n_{\mathbf{p}\sigma}$ of the momentum distribution function:
\begin{equation}
\delta(\frac{\Omega}{V})=\frac{1}{V}\sum_{\mathbf{p}\sigma}(\varepsilon_{\mathbf{p}\sigma}^0-\mu)\delta n_{\mathbf{p}\sigma}+\frac{1}{2V}\sum_{\mathbf{p}\sigma,\mathbf{p}'\sigma '}f_{\mathbf{pp'}}^{\sigma\sigma '}\delta n_{\mathbf{p}\sigma}\delta n_{\mathbf{p}'\sigma '}+...
\end{equation}
where $\varepsilon_{\mathbf{p}\sigma}^0$ is the quasi-particle energy, $f_{\mathbf{pp'}}^{\sigma\sigma '}$ are the quasi-particle interactions in the presence of the internal field. In the limit of a weakly ordered system, we can treat the quasi-particle interaction as rotationally invariant in spin space \cite{BS}, then
\begin{equation}
f_{\mathbf{pp'}}^{\sigma\sigma '}=f_{\mathbf{pp'}}^s+f_{\mathbf{pp'}}^a\sigma\cdot\sigma '+O(m_1^2)
\end{equation}

In 2-D, we start with the model:
\begin{equation}
{\bf m}_{\bf p}^{0}({\bf r})=-\frac{1}{N(0)}\frac{\partial n_{{\bf p}}^{0}}{\partial\varepsilon_{{\bf p}}^0}m_1(\hat{{\bf z}}\times\hat{{\bf p}})
\end{equation}
This model defines a spin orbit magnetism(SOM) state with zero net magnetization but non-zero spin current proportional to $m_1$, which can be seen as:
\begin{eqnarray}
{\bf \sigma}^0({\bf r})=2\sum{{\bf m}_{\bf p}^0}({\bf r})=0
\\
\begin{split}
{\bf j}_{{\bf\sigma},{\bf i}}&=2\sum_{\bf p}{v_{\bf p,i}{\bf m}_{\bf p}^0({\bf r})}(1+\frac{F_1^a}{2})
\\&=\frac{1}{2}v_f(1+\frac{F_1^a}{2})m_1({\bf\hat z\times{\bf\hat i}})
\end{split}
\end{eqnarray}

To understand the instability to this ground state, we first use Eq.(1) to calculate the free energy change based on this model using $[\delta n_\mathbf{p}]={\bf m}_{\bf p}^{0}\cdot\vec\sigma$ in spin space:
\begin{equation}
\delta(\frac{\Omega}{V})=\frac{1}{N(0)}(1+\frac{F_1^a}{2})m_1^2+\beta m_1^4
\end{equation}
which means that this ground state is protected by a generalized Pomeranchuck condition in the $F_1^a$ channel since we work in the ordered state. Here, $\beta>0$ is a phenomenological parameter making sure the model is valid and this term is the next leading order added by hand based on symmetry. The minimum of the free energy for $F_1^a<-2$ leads to the equilibrium order parameter(spin current) $m_1\sim \left|1+\frac{F_1^a}{2}\right|^{\frac{1}{2}}$, and in the limit $F_1^a\rightarrow -2^-$, $m_1$ is small, i.e. in the weakly ordered limit.

To study the collective modes around this new exotic magnetic Fermi liquid(EMFL) ground state, we investigate the free oscillation of the momentum dependent magnetization $\delta {\bf m_p}$. These oscillations of $\delta {\bf m_p}$ can be determined from the linearized Landau kinetic equation in the spin channel \cite{FL}:
\begin{widetext}
\begin{equation}
\frac{\partial {\delta\bf m_p}({\bf r},t)}{\partial t}+{\bf v_p}\cdot {\bf\nabla}(\delta{\bf m_p}({\bf r},t)-\frac{\partial n_{\bf p}^0}{\partial\varepsilon_{\bf p}^0}\delta{\bf h_p}({\bf r},t))\\=-2({\bf m_p}^0({\bf r},t)\times \delta{\bf h_p}({\bf r},t)+\delta{\bf m_p}({\bf r},t)\times{\bf h_p}^0({\bf r},t))+I[{\bf m_p}]
\end{equation}
where ${\bf h_p^0}=-{\bf B}+2\sum_{\bf p'}f_{\bf p p'}^a {\bf m_p'}$ and $\delta{\bf h_p}=-\delta{\bf B}+2\sum_{\bf p'}f_{\bf p p'}^a \delta{\bf m_p'}$ are the effective equilibrium field and its fluctuation, respectively. To study the free oscillations when ${\bf B}=0$ we set ${\delta\bf B}=0$. At low temperature the  the collision integral $I[{\bf m_p}]$ is negligible and it can be ignored in what follows.

To derive the dispersion relations, we do a Fourier transformation of Eq.(7), and plug in our model Eq.(3), and set $\delta{\bf m_p (q)}=(-\frac{1}{N(0)})\frac{\partial n_{\bf p}^0}{\partial\varepsilon_{\bf p}^0}{\vec{\nu}_{\bf p}({\bf q})}=\sum_{l}(-\frac{1}{N(0)})\frac{\partial n_{\bf p}^0}{\partial\varepsilon_{\bf p}^0}{\vec{\nu}}_{l}({\bf q})e^{il\phi_{\bf p}}$. Finally, Eq.(7) becomes:
\begin{equation}
\sum_{l,m}[\omega-{\bf q\cdot v_p}(1+\frac{F_{\left|l\right|}^a}{a_l})]{\vec{\nu}}_{l}({\bf q})e^{il\phi_{\bf p}}=2m_1 i\sum_{l,m}(\frac{f_{1}^a}{2}-\frac{f_{\left|l\right|}^a}{a_l})({\bf\hat z}\times{\bf\hat p})\times{\vec{\nu}_{l}({\bf q})}e^{il\phi_{\bf p}}
\end{equation}
where $a_l=\delta_{l,0}+2(1-\delta_{l,0})$.
\end{widetext}

Projecting Eq.(8) to each component of $e^{il\phi_{\bf p}}$, we take $l=0,1,-1$ component of the equation, and keep the expansion of $F_l^a$ only up to the $l=1$ term. The equations for the $l=0$, $l=1$ and $l=-1$ moment are:
\begin{eqnarray}
\omega{\vec{\nu}}_{0}=\frac{q v_f}{2}(1+\frac{F_1^a}{2})e^{i\phi_{\bf q}}{\vec{\nu}}_{1}+\frac{q v_f}{2}(1+\frac{F_1^a}{2})e^{-i\phi_{\bf q}}{\vec{\nu}}_{-1}\\
\omega{\vec{\nu}}_{1}=\frac{q v_f}{2}(1+F_0^a)e^{-il\phi_{\bf q}}\vec\nu_0+m_1i(f_0^a-\frac{f_1^a}{2})\vec\nu_0\times{\bf L}_1\\
\omega{\vec{\nu}}_{-1}=\frac{q v_f}{2}(1+F_0^a)e^{il\phi_{\bf q}}\vec\nu_0+m_1i(f_0^a-\frac{f_1^a}{2})\vec\nu_0\times{\bf L}_2
\end{eqnarray}
where ${\bf L}_1=(i,1,0)$ and ${\bf L}_2=(-i,1,0)$ are two complex vectors.
Considering each component of the vectors, we can solve these nine equations and get the dispersion relation of the collective modes. The dispersion relations for the gapless modes are given by:
\begin{equation}
\begin{split}
&\omega_c=\\
&\pm\frac{1}{2}\sqrt{\left|2+F_1^a\right|(2f_0^a-f_1^a)m_1 v_f q-\left|2+F_1^a\right|(1+F_0^a)v_f^2q^2}\\
&\rightarrow\pm\frac{1}{2}\sqrt{\left|2+F_1^a\right|(2f_0^a-f_1^a)m_1 v_f q}
\end{split}
\end{equation}
In this hydrodynamic-like approach, the truncation of the Fermi surface distortions up to $l=1$ is reasonable, since if we include the $l=2$ distortion terms, we will find that $\frac{\left|\vec\nu_{\pm 2}\right|}{\left|\vec\nu_{\pm 1}\right|}=\frac{qv_f}{2\omega}(1+\frac{F_1^a}{2})$, which is very small for small momentum transfer. In this sense, the inclusion of $\vec\nu_{\pm 2}$ will not qualitatively change the dispersion of the collective modes.

To determine if the mode in this EMFL is propagating or Landau damped, we need to consider the particle-hole(p-h) continuum. The continuum can be determined from Eq.(7) and for 2D we find that $\omega_{ph}^{\pm}={\bf q\cdot v_p}\pm m_1\left|f_1^a\right|$.

This mode is very exotic since it propagates with a $\sqrt q$ dispersion relation for small momentum unlike the magnons found in the Ferromagnetic and Antiferromagnetic phase. We realize that, due to the $\sqrt q$ dispersion, this mode will have higher order temperature dependence in, e.g., the specific heat, making it difficult to be detected in low temperature specific heat measurements. Given that it is separated from the p-h continuum, it may be possible using neutron scattering to detect this spin mode. Taking reasonable values of the Landau parameters and the order parameter, we evaluate the dispersion relation of the collective mode and p-h continuum. The result is presented in Fig.1.

In Fig.1, we have shown the collective mode together with the p-h continuum. Clearly, we can see that this gapless mode can propagate for small momentum and merges into the continuum for relatively large momentum.

\begin{figure}
  \centering
  \includegraphics[scale=0.8]{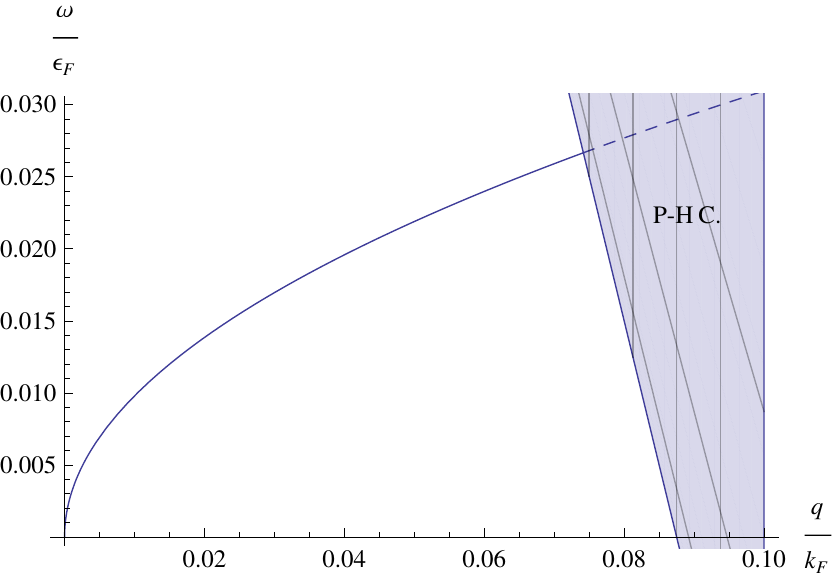}\\
  \caption{Collective mode together with the p-h continuum in 2D system. Here, we take $F_0^a=0.1$, $F_1^a=-2.1$, $m_1=0.12n$}\label{fig:figure1}
\end{figure}

We can check the validity of the hydrodynamic approach, for studying the collective modes, by calculating the dynamical spin response function using $\vec\chi=-\frac{\vec\nu_0}{\delta B}$. Here we use the Landau kinetic equation(Eq.7) \cite{PN} where we keep $\delta\mathbf{B}$ in the equation. By solving for the poles of the spin response function, we can also get the dispersion of the collective mode:
\begin{equation}
\omega_c=\pm\frac{1}{2}\sqrt[4]{1-\frac{f_0^a}{2f_0^a-f_1^a}}\sqrt{\left|2+F_1^a\right|(2f_0^a-f_1^a)m_1v_fq}
\end{equation}
which is consistent with the result we found in the previous hydrodynamic-like approach with $\sqrt{q}$ dispersion in leading order. In comparing Eq.(12) and Eq.(13), the leading order behavior is not exactly the same. This is due to the fact that in the previous hydrodynamic-like approach, we truncated the Fermi surface distortion at $l=1$. In the calculation of the response function, we truncate the Landau parameters at $l=1$, but we keep the Fermi surface distortion to all order. Although, the inclusion of the higher order distortions won't dramatically change the leading order $\sqrt{q}$ behavior, which is already shown above, it can still slightly modify the prefactor.

In this EMFL state it is possible that the new spin wave mode could give rise to a superconducting instability. The response function for this mode for small momentum and energy transfer is approximately given by:
\begin{equation}
\chi\sim\frac{2N(0)}{F_0^a-\frac{F_1^a}{2}}\frac{\omega_c^2}{\omega^2-\omega_c^2}
\end{equation}
The structure resembles that of response function of a phonon, which makes it possible that the spin fluctuation mediated interaction can cause the pairing of two quasi-particles, and lead to superconductivity. Since this pairing is caused by spin fluctuations, we expect that the superconductivity is unconventional, in the sense that the pairing symmetry is different from the normal s-wave phonon mediated superconductors. It's actually p-wave, which is demonstrated below by the argument from the forward scattering sum rule.

Within the framework of Landau Fermi liquid theory, based on the forward scattering sum rule \cite{FL,BEB}, we can demonstrate the instability towards superconductivity and analyze the pairing symmetry of it. In Fermi liquid theory, the scattering amplitude for small momentum transfer can be expanded as $N(0)a_{\mathbf{p} \mathbf{p'}}^{\sigma\sigma'}=\sum_l{(A_l^s+A_l^a \sigma\sigma')}P_l(\mathbf{\hat p}\cdot\mathbf{\hat p'})$\cite{FL}. In the case of weak magnetic ordering the quasi-particle scattering amplitude, $A_l^\alpha$ can be expressed by Landau parameters as $A_l^\alpha=\frac{F_l^\alpha}{1+F_l^\alpha/a_l}$, where $\alpha=a,s$ and $a_l$ has the same definition as above \cite{FL}. The forward scattering sum rule states that the triplet scattering of two quasi-particles with the same momenta must vanish. Therefore, to the leading order of $m_1$, we have $\sum_l(A_l^s+A_l^a)=0$. Since in our model, we only consider the interaction up to $l=1$, we can truncate the equation up to $l=1$, then
\begin{equation}
A_0^a+A_0^s+A_1^a+A_1^s=0
\end{equation}
In our magnetically ordered state close to the phase transition, $F_1^a\rightarrow -2^-$, and it follows that $A_1^a\rightarrow+\infty$, which requires at least one of the first three terms in Eq.(15) to diverge as $-\infty$ when approaching the transition point. Firstly, the diverging of $A_1^s$ implies the vanishing of the effective mass, and since we assume a finite density of state on the Fermi surface, it won't occur in our system. Then only $A_0^s$ and $A_0^a$ are left to satisfy Eq.(15). Taking $A_0^s$ as an example, let $A_0^s\rightarrow -A_1^a$ diverge to $-\infty$, which indicates instabilities in both spin and charge sectors respectively. This leads to phase separation at the point of the magnetic phase transition. We can now look at the scattering amplitude in both spin singlet and triplet channels, where the expansion is still truncated up to l=1:
\begin{eqnarray}
a_{\mathbf{p}\mathbf{p'}}^{singlet}=A_0^s-3A_0^a+(A_1^s-3A_1^a)(\mathbf{\hat p}\cdot\mathbf{\hat p'})
\\
a_{\mathbf{p}\mathbf{p'}}^{triplet}=A_0^s+A_0^a+(A_1^s+A_1^a)(\mathbf{\hat p}\cdot\mathbf{\hat p'})
\end{eqnarray}
In the magnetically ordered state close to the transition, consider the scattering of a pair of quasi-particles with opposite momentum, the scattering amplitude becomes:
\begin{eqnarray}
a^{singlet}=2A_1^a\rightarrow+\infty
\\
a^{triplet}=-2A_1^a\rightarrow-\infty
\end{eqnarray}
Obviously, we see a strong repulsion in the singlet channel and a strong attraction in the triplet channel, indicating an instability towards p-wave superconductivity. The same scenario happens if we let $A_0^a$ diverge.

We also study the same model in a 3D system, where the Fermi surface distortion is very different from that in the 2D case. In the 2D system, since the quasi-particle momentum $\bf p$ lives in the xy plane, the magnitude of ${\bf m_p}$ is independent of the direction of $\bf p$, which means the Fermi surface distortion is isotropic and there is a constant gap between the two branches of the Fermi surface with different spin polarization. In a 3D system, however, the gap will depend on the direction of $\bf p$ and there are nodes located at the north and south pole of the Fermi surface which makes the p-h continuum very different from that in the 2D case. The p-h continuum is no longer gapped at zero momentum, instead, it sweeps a finite region at zero momentum, which will Landau damp the $\sqrt q$ mode and it will not propagate at all. In order to avoid this problem, we introduce an additional Ferromagnetic order in our model, which will gap out the p-h continuum at zero momentum so that a small window will be opened to let the collective mode propagate. So the new model becomes:
\begin{equation}
{\bf m}_{\bf p}^{0}({\bf r})=-\frac{1}{N(0)}\frac{\partial n_{{\bf p}}^{0}}{\partial \varepsilon_{{\bf p}}^{0}}[m_{0}\hat{{\bf z}}+m_{1}(\hat{{\bf z}}\times\hat{{\bf p}})]
\end{equation}
which defines a state with magnetization proportional to $m_0$ and spin current proportional to $m_1$, similar to the 2D case except for the non-zero magnetization.

Similarly to what we do in the 2D case, we can also calculate the free energy change based on this model:
\begin{equation}
\delta(\frac{\Omega}{V})=\frac{1+F_0^a}{N(0)}m_0^2+\frac{2}{3N(0)}(1+\frac{F_1^a}{3})m_1^2+o(m_0^2,m_1^2)
\end{equation}
which means that this ground state is also protected by generalized Pomeranchuck conditons. Since there are multiple order parameters, it is necessary to study the competition between the different order parameters. Using the same hydrodynamic-like approach as in the 2D case, we find two modes in the phase where both orders survive:
\begin{eqnarray}
\omega_1\rightarrow 2D_0+(\frac{A_0A_1}{6D_0}+\frac{A_1^2D_1^2}{36D_0^3})q_\bot^2
\\
\omega_2\rightarrow\frac{2\left|A_1\right|\sqrt{q-\left|\frac{2D_1}{A_0}\right|}}{\sqrt{\frac{9A_0D_0^2+6A_1D_1^2}{D_1^3}}}
\end{eqnarray}
where,
\[A_0=(1+F_0^a)v_f,A_1=(1+\frac{F_1^a}{3})v_f\]
\[D_0=m_0(f_0^a-\frac{f_1^a}{3}),D_1=m_1(f_0^a-\frac{f_1^a}{3})\]
and the p-h continuum is:
\[\omega_{ph}={\bf q\cdot v_p}\pm\frac{2}{3}\sqrt{9(m_0f_0^a)^2+(m_1f_1^a)^2\sin^2(\theta_p)}\]
Here, since $m_0>0$, the p-h continuum is gapped, which opens up a window for the modes to propagate. We evaluate the collective modes and p-h continuum with reasonable values of Landau parameters and order parameters, and the result is presented in Fig.2.

\begin{figure}[ht]
\begin{center}
\begin{tabular}{cccccc}
\includegraphics[scale=0.6]{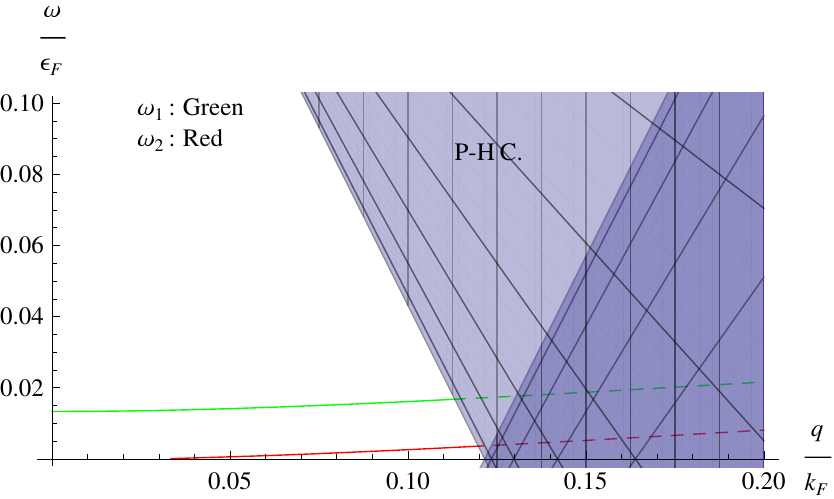}\\
		(a)\\
\includegraphics[scale=0.6]{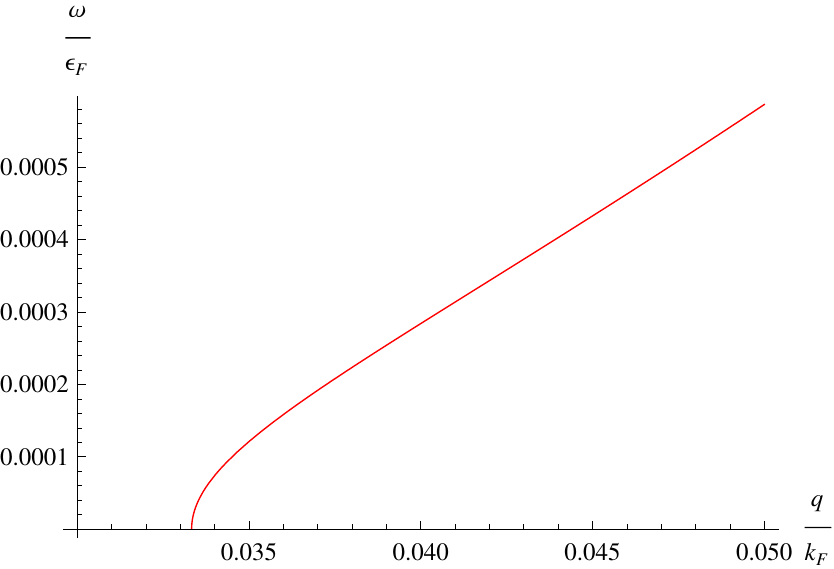}\\
        (b)\\
\end{tabular}
\end{center}
\caption{(a)Dispersions of the collective modes with the p-h continuum in phase where both order coexist. (b)Zoom in version of mode $\omega_2$. Here, we take $F_0^a=-1.1$, $F_1^a=-3.1$, $m_0=0.15n$, $m_1=0.075n$}\label{fig:figure2}
\end{figure}

In Fig.2, we have shown the gapless and gapped modes outside the p-h continuum. Clearly, we can see that $\omega_2^2<0$ for small q, which is a very exotic feature. This indicates a Lifshitz-like instability \cite{POM} of the ground state towards some inhomogeneous magnetic state such as a spiral phase \cite{L}.

In summary, using Landau Fermi liquid theory, we studied the collective modes in the spin orbit order magnetic state in the $f_1^a$ channel, in both 2D and 3D systems. In both cases, the $\sqrt{q}$ dispersion is found in leading order. In the 2D system, we also calculate the spin density response function, which gives a consistent result($\sqrt{q}$ dispersion) for the collective mode, suggesting that the hydrodynamic description captures the essential physics of the state. This exotic mode can play a role in the formation of cooper pairs of two quasi-particles since it has similar structure to the phonon propagator, so we expect an instability toward superconductivity close to the magnetic phase transition. A further argument based on forward scattering sum rules confirms the instability again and strongly indicates a p-wave pairing symmetry. In a 2D system, the model describes one general structure of spin-orbital coupling and it's actually closely related the Rashba Hamiltonian \cite{R} in the 2D semiconductor heterostructures. Therefore, we expect that this model can describe 2D or quasi-2D systems with spin-orbital coupling. In 3D, a Ferromagnetic order is added to the ground state to avoid the Landau damping and the collective mode leads to a Lifshitz-like instability towards an inhomogeneous magnetic state in one of the phases.

We would like to thank Sasha Balatsky and Jan Engelbrecht for valuable discussion and advice. Also KB would like to thank Jason Jackiewicz for valuable discussions during the early phase of the project.

\end{document}